\documentclass[12pt]{article}
\usepackage{graphicx}
\usepackage{dcolumn}
\usepackage{bm}
\usepackage{latexsym}

\begin{document}

\begin{center}
{\large\bf Photon-assisted electron transport through a three-terminal quantum dot system with nonresonant tunneling channels}

\vspace{0.5cm}

\medskip
T. Kwapi\'nski\footnote{e-mail: kwapin@hektor.umcs.lublin.pl}, R. Taranko \footnote{e-mail: taranko@tytan.umcs.lublin.pl}, E.
Taranko \footnote{e-mail: ewatara@tytan.umcs.lublin.pl}
\\
\vspace{0.5cm}

Institute of Physics, Maria Curie-Sk\l odowska University, \\
and\\
Maria Curie-Sk\l odowska University Nanotechnology Center \\
 20-031 Lublin, Poland
\end{center}

\vspace{0.5cm}

\begin{abstract}
We have studied the electron transport through a quantum dot coupled to three leads in the presence of external microwave fields
supplied to different parts of the considered mesoscopic system. Additionally,  we introduced a possible nonresonant tunneling
channels between leads. The quantum dot charge  and currents were determined in terms of the appropriate evolution operator
matrix elements and under the wide band limit the analytical formulas for time-averaged currents and differential conductance
were obtained. We have also examined the response of the considered system on the rectangular-pulse modulation imposed on
different quantum dot-leads barriers as well as the time-dependence of currents flowing in response to suddenly removed (or
included) connection of a quantum dot with one of the leads.
\end{abstract}

\section{Introduction}
The electron transport via resonant tunnelling in  mesoscopic systems has been the subject of extensive theoretical research due
to recent development in fabrication of small electronic devices and their potential applications. Some interest has been
focused on the transport properties of a quantum dot (QD) under the influence of external time-dependent fields. New effects
have been observed and theoretically described, e.g., photon-assisted tunnelling through small quantum dots, \cite{1,2},
photon-electron pumps \cite{2} and others. In most theoretical investigations a QD placed between two leads was considered
(e.g., Refs. \cite{2,4,5,6,7,8,29b}) and the current flowing through a QD under periodic modulation of the QD electronic
structure or periodic (nonperiodic) modulation of the tunnelling barriers  and electron energy levels of both (left and right)
leads was calculated.

One of the important problems of the mesoscopic physics is the interference of the charge carries. This interference appears
when two (or more) transmission channels for electron tunnelling  exist. Such possibility exists, e.g. in the electron transport
through a QD embedded in a ring in the Aharonov-Bohm geometry and much theoretical interest has been paid to description of the
phase coherence in this and related systems, e.g. Refs. \cite{9,10,11}. Another experimental situation in which the interference
may occur can be realized in the scanning tunnelling microscope (STM). The recent experimental and theoretical studies with a
low-temperature STM of a single atom deposited on a metallic surface showed the asymmetric Fano resonances in the tunnelling
spectra, e.g. Refs. \cite{6,12,13}. In the STM measurements tip probes the transmission of electrons either through the adsorbed
atom or directly from the surface. The transport of electrons through both channels leads to an asymmetric shape of the
conductance curves which is typical behavior for the Fano resonance resulting from constructive and destructive interference
processes. The quantum interference can be also observed in the mesoscopic system with multiple energy levels \cite{14}. A model
which incorporates a weak direct nonresonant transmission through a QD as well as the resonant tunneling channel was also
considered in Ref. \cite{15} in the context of the large value of the transmission phase found in the experiment for the Kondo
regime of a QD \cite{16}.

A number of works has been devoted to the problem of the electron transport in the multiterminal QD systems and here we mention
only a few of them. In Ref. \cite{17} the conductance of the N-lead system was considered showing that the Kondo resonance at
equilibrium is split into N-1 peaks. In Ref. \cite{18} an explicit form for the transmission coefficient in the electron
transport through a QD connected with three leads was found. The electron transport and shot noise in a multi-terminal coupled
QD system in which each lead was disturbed by classical microwave fields were studied in Ref. \cite{19}. Multiterminal QD
systems or magnetic junctions were also intensively investigated in  context of the spin-dependent transport, e.g. Refs.
\cite{20,21}. A three-terminal QD system was studied in Ref. \cite{21a,21b} to measure of the nonequilibrium QD density of
states (splitting of the Kondo resonance peak). The cross correlations of the currents and the differential conductance of the
QD coupled with three leads described by the infinite-U Anderson Hamiltonian were considered in Ref. \cite{21c}. The general
formulation of the time-dependent spin-polarized transport in a system consisting of the resonant tunnelling structure coupled
with several magnetic terminals was considered by Zhao \emph{et al}. \cite{21} and as an application of this formalism the
electron transport in a system with two terminals under an ac external field was investigated.
\begin{figure}[tb]
\begin{center}
\includegraphics[width=0.6\columnwidth]{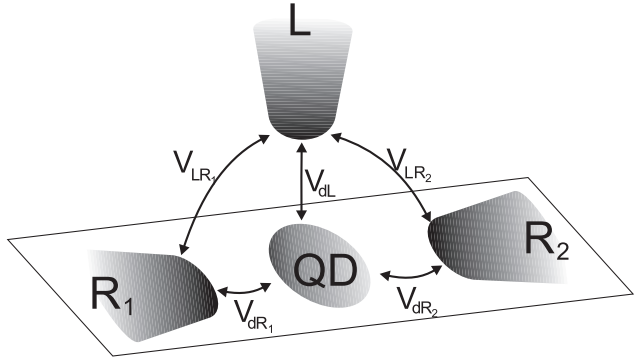}
\end{center}
\caption{Schematic picture of the QD coupled with three leads. It can serve as a possible STM experimental setup when the left
lead (L) corresponds to the STM tip.} \label{fig1}
\end{figure}

In all papers mentioned above and in the studies relating to the electron transport through a QD with the additional (bridge)
nonresonant transmission channels, the time-dependent external fields were not applied and the considered systems were driven
out of the equilibrium only by means of a dc voltage bias (see, however, Ref. \cite{26})

In this paper, we address the issue of a QD coupled with three leads with additional, nonresonant coupling between leads driven
out of equilibrium by means of a dc voltage bias and time-dependent external fields. The QD is connected with three metal leads
and one of these leads, say the left (L) lead, is coupled with the two remaining leads, say the first ($R_1$) and the second
($R_2$) right leads. The possible STM experimental setup corresponding to our model system is presented in Fig. 1. In
literature, different theoretical approaches have been developed to treat the time-dependent electron transport in the
mesoscopic systems. The most popular nonequilibrium Green's function method depends on the two time arguments and for
time-dependent problems it is a rather difficult task to calculate them without any approximations (e.g. beyond the wide-band
limit). Therefore, in our treatment of the time-dependent problems we use the evolution operator which, as a rule, essentially
depends on one time argument (e.g. Refs. \cite{22,23,24}). Such an approach is especially well suited for the problem with
time-dependent coupling between the QD and leads.

The outline of the paper is as follows. In Sec. II we start with the model and method for the derivation of the QD charge and
currents. In Sec. III we present the results for the time-averaged currents and their derivatives with respect to the QD energy
level position (or equivalently, with respect to the gate voltage) obtained for different time-dependence of the parameters
characterizing the considered system. We consider also the transient current characteristics in the case of the
rectangular-pulse modulations imposed on the QD-lead barriers. The last section presents the conclusions and in the Appendix we
give the short derivation of the evolution operator matrix elements needed in the QD charge and current calculations.

\section{Model and formalism}

We consider a QD coupled through the tunneling barriers $V_{\vec k_id}$ $(i=1,2,3)$ to three metal leads. One of these leads,
say the left lead (L) is coupled additionally with the remaining two leads, say the first and second right leads ($R_1$, $R_2$)
by the tunneling barriers $V_{\vec k_L, \vec k_R}$. In the following we will denote the wave vectors associated with the left
lead by the letter $\vec k$ and the wave vectors corresponding to the first and second right leads by the letters $\vec q$ and
$\vec r$, respectively. The chemical potentials $\mu_i$ of the three metal leads may not be equal, and their difference is not
necessarily small. We write the Hamiltonian of the considered system in the form $H = H_0 + V$, where
\begin{equation}
 H_0 = \sum_{\vec p=\vec k,\vec r,\vec q} \varepsilon_{\vec p}(t) a^+_{\vec p}
 a_{\vec p} + \varepsilon_d(t) a^+_d a_d\,,
\end{equation}
\begin{eqnarray}
 V & = &\sum_{\vec p=\vec k,\vec r,\vec q}(V_{\vec pd}(t) a^+_{\vec p}a_d + {\rm h.c.}) +\nonumber\\
&+& \sum_{\vec k,\vec r} (V_{\vec k\vec r}(t)a^+_{\vec k}a_{\vec r} +
 {\rm h.c.}) + \sum_{\vec k,\vec q} (V_{\vec k\vec q}(t)a^+_{\vec k}a_{\vec q} + {\rm h.c.})
\end{eqnarray}
The operators $a_{\vec p}(a^+_{\vec p})$, $a_d(a^+_d)$ are the annihilation (creation) operators of the electrons in the
corresponding leads and the dot, respectively. For simplicity the dot is characterised only by a single level $\varepsilon_d$
and the intra-dot electron-electron Coulomb interaction is neglected. The neglect of Coulomb interaction is quite reasonable in
some systems and, as we are going to concentrate on the investigations of the influence of the third lead (in comaprison with
the QD-two leads system) and the additional tunneling channels between the leads on the time-dependent transport, then in the
first step ignoring  the Coulomb interaction should be justified. We consider our mesoscopic system in the presence of external
microwave (MW) fields which are applied to the dot and three leads. In most theoretical treatments of photon-assisted electron
tunneling it is assumed that the driving field equals the applied external field. However, the situation is more complicated and
the internal potential can be different from the applied potential \cite{27}. One of the consequences will be, e.g. the
asymmetry between the corresponding left and right sidebands \cite{25,26}. The main feature of the time-dependent transport
remains, however, unchanged and in our treatment as usual we assume that in the adiabatic approximation the energy levels of the
leads and QD are driven with the frequency $\omega$ and the amplitudes $\Delta_i$ ($i=L, R_1, R_2$), $\Delta_d$ and read
$\varepsilon_{\vec k_i}(t)=\varepsilon_{\vec k_i} + \Delta_i\cos\omega t$, $\varepsilon_d(t) = \varepsilon_d +
\Delta_d\cos\omega t$, respectively.

The time-evolution of the QD charge and the current flowing in the system can be described in terms of the time-evolution
operator $U(t,t_0)$  given by the equation of motion (in the interaction representation):
\begin{eqnarray}
&& i{\partial U(t,t_0)\over \partial t} = \tilde V(t) U(t,t_0) ,
\end{eqnarray}
with $\tilde V(t) = U_0(t,t_0) \, V(t) U^+_0(t,t_0)$ and $U_0(t,t_0) = T {\rm
exp}\left(i\int\limits^t_{t_0}dt_1\,H_0(t_1)\right)$ where $T$ denotes the time ordering and the units such that $\hbar=1$ have
been chosen. Here we have assumed that the interactions between the QD and leads, as well as between the left and right leads
are switched on in the distant past $t_0$.

The QD charge is calculated according to the formula (cf. Refs. \cite{22,23}):
\begin{equation}
 n_d(t) = n_d(t_0)|U_{dd}(t,t_0)|^2 + \sum_{\vec p=\vec k,\vec r,\vec q}
 n_{\vec p} (t_0)|U_{d\vec p}(t,t_0)|^2 .
\end{equation}
Here $U_{dd}(t,t_0)$ and $U_{d\vec p}(t,t_0)$ denote the matrix elements of the evolution operator calculated within the basis
functions containing the single-particle functions $|\vec k\rangle$, $|\vec q\rangle$, $|\vec r\rangle$ and $|d\rangle$
corresponding to three leads and QD, respectively. $n_d(t_0)$ and $n_{\vec p}(t_0)$ represent the initial filling of the
corresponding single-particle states.

The tunneling current flowing, e.g. from the left lead, $j_L(t)$, can be obtained using the time derivative of the total number
of electrons in the left lead, $j_L(t) = -edn_L(t)/dt$ (in the following we put $e = 1$), where
\begin{eqnarray}
 n_L(t) &=& \sum_{\vec k} n_{\vec k}(t) = \sum_{\vec k}  n_d(t_0)
   |U_{\vec k d}(t,t_0)|^2 + \nonumber\\
 &+& \sum_{\vec k,\vec k_1} n_{\vec k_1}(t_0)|U_{\vec k\vec k_1}
  (t,t_0)|^2 + \nonumber \\
 &+& \sum_{\vec k, \vec q} n_{\vec q}(t_0)|U_{\vec k\vec q}(t,t_0)|^2
 +\sum_{\vec k, \vec r} n_{\vec r}(t_0)|U_{\vec k\vec r}(t,t_0)|^2\,.
\end{eqnarray}
In the following only the matrix elements of the evolution operator present in Eqs. (4) and (5) are required and they can be
obtained solving the corresponding sets of coupled differential equations constructed according to Eq. (3) with $\tilde
V_{ab}(t)$ written as follows:
\begin{equation}
 \tilde V_{ab}(t)=V_{ab}(t) {\rm exp} \left(i(\varepsilon_a - \varepsilon_b)(t - t_0)
 + i{\Delta_a - \Delta_b\over \omega} (\sin\omega t -\sin\omega t_0)\right) \,,
\end{equation}
where $a$ and $b$ correspond to $|\vec k\rangle$, $\vec q\rangle$, $|\vec r\rangle$ or $|d\rangle$, respectively.

As the example, the matrix element $U_{dd}(t,t_0)$ required for the calculation of the first term of the QD charge, Eq. 4, can
be obtained solving the following set of coupled differential equations
\begin{eqnarray}
&& {\partial U_{dd}(t,t_0)\over\partial t} = -i\sum_{\vec p=\vec k,\vec q,\vec r}
 \, \tilde V_{d\vec p} (t) U_{\vec p d}(t,t_0)\,, \\
&& {\partial U_{\vec kd}(t,t_0)\over\partial t} = -i\tilde V_{\vec kd}(t)
 U_{dd}(t,t_0) - i\sum_{\vec p=\vec q,\vec r} \, \tilde V_{\vec k \vec p}(t)\,
 U_{\vec pd}(t,t_0)\,, \\
&& {\partial U_{\vec pd}(t,t_0)\over\partial t} = -i\tilde V_{\vec pd}(t)
 U_{dd}(t,t_0) - i\sum_{\vec k} \, \tilde V_{\vec p \vec k}(t)\,
 U_{\vec kd}(t,t_0)\, , \,\,\, \vec p=\vec q, \vec r \,.
\end{eqnarray}
The total number of coupled equations in this case is equal to $(3N+1)$, $N$ being the number of discrete wave vectors $\vec k$,
$\vec q$ and $\vec r$ taken to perform the corresponding summation over the wave vectors. Usually, the number $N$ equal to
100-200 is sufficient to achieve the desired accuracy of the calculations. We have solved numerically this and other similar
sets of the coupled differential equations needed in calculations of all matrix elements of the evolution operator present in
Eqs. 4,5. We have used this method for the special case of time-dependent couplings of the QD with leads and the couplings of
the left lead with two right leads. The set of Eqs. 7-9 in the case of vanishing over-dot couplings between the left and right
leads is greatly simplified and gives, e.g. for $U_{dd}(t,t_0)$:
\begin{eqnarray}
 {\partial U_{dd}(t,t_0)\over\partial t} = -\int_{t_0}^t dt_1  \texttt{K}(t,t_1)U_{dd}(t,t_0)\,,
\end{eqnarray}
where
\begin{eqnarray}
\texttt{K}(t,t_1)&=&\sum_{\vec p=\vec k,\vec q,\vec r} \tilde V_{d\vec p} (t) \tilde V_{\vec p d} (t_1)= \nonumber\\
&& \sum_{i=L, R_1, R_2} |V_i|^2 u_i(t) u_i(t_1) \texttt{D}_i(t-t_1) \exp(i \varepsilon_d (t-t_1))  \nonumber\\
&&\times \exp(i(\Delta_d-\Delta_i)(\sin \omega t-\sin \omega t_1)/\omega)
\end{eqnarray}
and $\texttt{D}_i(t-t_1)$ is the Fourier transform of the $i$-th lead density of states and $V_{di}(t)=V_i u_i(t)$. Similar
simplifications occur in the calculations of other matrix elements of $U(t,t_0)$ required in the formulas for $n_d(t)$ and
$n_L(t)$. However, for the nonvanishing couplings $V_{\vec k \vec q}$ and $V_{\vec k \vec r}$ (over-dot bridge between the left
and right leads) one has to solve the starting set of Eqs. 7-9. Much more analytical calculations can be done in the case of
constant values of the tunneling matrix elements present in our model. In this case the general equation satisfied by
$U_{dd}(t,t_0)$ is derived in the Appendix and under the wide band limit (WBL) approximation, e.g. \cite{2,4,5} this equation
takes the simple form
\begin{equation}
  {\partial U_{dd}(t,t_0)\over \partial t} = -C_1\, U_{dd}(t,t_0) ,
\end{equation}
here $ C_1 = \left( {3 \over 2} - {3x^2 + 2ix \over 1 + 2x^2}\right) \Gamma$, $x = \pi V_{RL}/D$, $D$ being the bandwidth of the
lead energy band ($D_{R_1}=D_{R_2}=D_{L}=D$) and $\Gamma=2 \pi V^2/D$. In the Appendix we give the derivations of all functions
needed for calculation of the QD charge and currents. We assumed the simplified assumption that all tunneling matrix elements
are independent of the wave vectors. The interactions between the QD and leads are assumed to be equal between themselves and
denoted by $V$ and the interactions between the left and two rights leads (i.e. $V_{\vec k\vec q}$ and $V_{\vec k\vec r}$)
corresponding to the over-dot tunneling channels are also equal one with another and denoted by $V_{RL}$.

It is easy to show that the first term of the general formula for the QD charge, Eq. 4, together with the solution of  Eq. (12),
$U_{dd}(t,t_0)=exp(-C_1(t-t_0))$, tends to zero for $t - t_0 \rightarrow \infty$ as ${\texttt{Re}}C_1={3 \Gamma \over 2}-{3x^2
\Gamma \over 1+2x^2} > 0$.  The next terms of the QD charge formula can be calculated using the functions $U_{d\vec k}(t,t_0)$,
$U_{d\vec q}(t,t_0)$ and $U_{d\vec r}(t,t_0)$, Eqs. 43,47, being the solutions of the corresponding differential equations, Eqs.
42,46. Finally, the time-averaged QD charge is given by
\begin{equation}
 \langle n_d(t) \rangle= \sum_{i=L,R_1,R_2} a_i \, \int d\varepsilon f_i(\varepsilon)
 \langle|A_i(\varepsilon,t)|^2\rangle\,,
\end{equation}
where
\begin{eqnarray}
&& a_L =  (1 + 4x^2)/(1 + 2x^2)^2 \Gamma/2\pi\,,\\
&& a_{R_1} = a_{R_2} = (1+x^2)/(1 + 2x^2)^2 \Gamma/2\pi\,.
\end{eqnarray}
\begin{eqnarray}
A_i(\varepsilon,t) &=& -i\int\limits^t_{t_0} dt_1 {\rm exp}
 \left( -i(\varepsilon_d - \varepsilon)(t - t_1) - i(\Delta_d - \Delta_i) \right.
 \nonumber\\ && \left. (\sin\omega t -  \sin\omega t_1)/\omega)\right) \, {\rm exp} \left({\Gamma (-3 + i4x) \over 2(1 + 2x^2)}
 (t - t_1)\right) \,,
\end{eqnarray}
where $\langle ...\rangle$ denotes the time-averaging and $f_i(\varepsilon)$ denotes the Fermi function of the $i$-th ($i = L,
R_1, R_2)$ lead. Noticing, that ${\texttt{Im}}\langle A_i(\varepsilon, t) \rangle = -{3 \Gamma/2 \over 1+2x^2} \langle
|A_i(\varepsilon, t)|^2 \rangle $ (cf. \cite{5}), the expression for the time-averaged QD charge can be written as:
\begin{eqnarray}
 \langle n_d(t)\rangle &=& -{\texttt{Im}}\left( {1+4x^2 \over 3 \pi (1+2x^2)} \int f_L(\varepsilon) \langle A_L(\varepsilon, t) \rangle
 d\varepsilon  \right. \nonumber\\ &+& \left.  {1+x^2 \over 3 \pi (1+2x^2)}  \sum_{i=1,2}  \int f_{R_i}(\varepsilon)
 \langle A_{R_i}(\varepsilon,t) \rangle\ d\varepsilon  \right) \,.
\end{eqnarray}
In order to calculate the current $j_L(t)$ the
 functions $U_{\vec kd}(t,t_0)$, $U_{\vec k_1\vec k_2}(t,t_0)$,
$U_{\vec k\vec q}(t,t_0)$ and $U_{\vec k\vec r}(t,t_0)$ are required and they are given in the Appendix in Eqs. 37, 44, 49.
After lengthly but straightforward calculations we obtain for the time averaged current leaving the left lead the following
formula:

\begin{eqnarray}
\langle j_L(t)\rangle &=& \sum_{i=R_1, R_2} {\texttt{Re}} \left[ {2x^2 \over \pi (1+2x^2)^2} (\mu_L-\mu_i) \right. \nonumber\\
&+& \left.  G \left( \int f_L(\varepsilon) \langle A_L(\varepsilon, t) \rangle d\varepsilon - \int f_i(\varepsilon) \langle
A_i(\varepsilon, t) \rangle d\varepsilon \right) \right] \,,
\end{eqnarray}
where
\begin{eqnarray}
G={\Gamma \over 3 \pi (1+2x^2)^3} \left(6x(1-2x^2)+i(1-13x^2+4x^4) \right)\,,
\end{eqnarray}
\begin{eqnarray}
\langle A_i(\varepsilon, t)\rangle = \sum_k J_k^2 \left( {\Delta_d-\Delta_i \over \omega}\right) \left(
\varepsilon-\varepsilon_d-\omega k+{2\Gamma x \over 1+2x^2}+i {3\Gamma/2 \over 1+2x^2} \right)^{-1} \,,
\end{eqnarray}
 and $J_k(y)$ denotes the Bessel
function. The corresponding formula for the time-averaged current $\langle j_{R_i}(t)\rangle$ leaving $R_i$-lead, $i=1,2$,
cannot be written in such symmetrical form as in Eq. 18, because the $R_i$-lead is coupled with $L$-lead only. For $\langle
j_{R_i}(t)\rangle$ we have:
\begin{eqnarray}
\langle j_{R_i}(t)\rangle &=&
{\texttt{Re}} \left[ {2x^2 \over \pi (1+x^2)^2} \left(\mu_{R_i}- \mu_L+x^2(\mu_{R_i}-\mu_{R_j})\right) \right. \nonumber\\
&+& {\Gamma \over 3\pi (1+2x^2)^3}  \left( 2 G_2 \int f_{R_i}(\varepsilon) \langle A_{R_i}(\varepsilon, t) \rangle d\varepsilon
-G_1 \int f_{L}(\varepsilon) \langle A_{L}(\varepsilon, t) \rangle d\varepsilon \right. \nonumber\\  &-& \left. \left. G_3 \int
f_{R_j}(\varepsilon) \langle A_{R_j}(\varepsilon, t) \rangle d\varepsilon \right) \right] \,,
\end{eqnarray}
where $G_1=6x(1-2x^2)+i(1-13x^2+4x^4)$, $G_2=3x-{i \over 2}(-2+5x^2+x^4)$, $G_3=12x^3+i(1+8x^2-5x^4)$ and $j=1(2)$ for $i=2(1)$.
Note, that the integrals present in the formula for the QD charge and currents, Eqs. 17,18,21, can be easily performed
analytically and final algebraic expressions can be obtained. Especially simple and transparent form can be given for the
conductance ${\partial \over \partial \mu_i} \langle j_j(t) \rangle$, $i,j=L,R_1, R_2$. For example, ${\partial \langle j_L(t)
\rangle \over \partial \mu_L}$ reads as:
\begin{eqnarray}
{\partial \over \partial \mu_L} \langle j_L(t)  \rangle &=& {4x^2 \over \pi (1+2x^2)^2} + \sum_k J_k^2 \left( {\Delta_d-\Delta_L
\over \omega}\right) \nonumber\\ && F_1 \left( {\Gamma^2 (1-13x^2+4x^4) \over \pi (1+2x^2)^4} + {4\Gamma x (1-2x^2)\over \pi
(1+2x^2)^3}F_2 \right)\,,
\end{eqnarray}
where
\begin{eqnarray}
F_1=\left[ \left(\mu_L-\varepsilon_d-\omega k + {2\Gamma x \over 1+2x^2}\right)^2+\left( {3\Gamma/2 \over 1+2x^2}\right)^2
\right]^{-1} \,,
\end{eqnarray}
\begin{eqnarray}
F_2=\left(\mu_L-\varepsilon_d-\omega k + {2\Gamma x \over 1+2x^2}\right) \,.
\end{eqnarray}
Analyzing Eq. 22 one can find the origin of the asymmetric line shapes in differential conductance resulting from the
interference of resonant and nonresonant tunneling paths. For the case of $V_{LR}=0$ we observe the Lorentzian resonances
localized at $\varepsilon_d=\mu_L \pm \omega k$. The amplitudes of these resonances are determined by the $k$-th order Bessel
functions calculated for the argument ${(\Delta_d-\Delta_i)/ \omega}$. For the case of nonvanishing $V_{LR}$, the resulting
curve is a superposition of the Lorentzian-like resonances and asymmetric parts weighed by the factors ${\Gamma^2 }
(1-13x^2+4x^4)/(1+2x^2)^4$ and $4\Gamma x (1-2x^2)/\pi(1+2x^2)^3$, respectively. The Lorentzian-like resonance is centered at
$\varepsilon_d=\mu_L \pm \omega k+2\Gamma x/(1+2x^2)$ with the peak width at half maximum (FWHM) equal to $3\Gamma/(1+2x^2)$ and
the maximum value equal to ${4\over 9\pi}{1-13x^2+4x^4 \over (1+2x^2)^2}J_k^2 \left( {\Delta_d-\Delta_L \over \omega} \right)$.
The asymmetric part of the differential conductance corresponding to the $k-$th sideband is also centered in the same point with
the distance between its maximum and minimum equal to ${3\Gamma \over 1+2x^2}$ and the absolute values of these extrema are
equal to ${4\over 3\pi}{x (1-2x^2) \over (1+2x^2)^2}J_k^2 \left( {\Delta_d-\Delta_L \over \omega} \right)$. For comparison, in
the case of the QD coupled with two leads the corresponding Lorentzian-like part of the differential conductance corresponding
to the $k-$th sideband is centered at $\varepsilon_d=\mu_L \pm \omega k+ \Gamma x/(1+x^2)$, with FWHM equal to $2\Gamma/(1+x^2)$
and the maximum value equal to ${1\over 2\pi}{1-6x^2+x^4 \over (1+2x^2)^2}J_k^2 \left( {\Delta_d-\Delta_L \over \omega}
\right)$. Knowing the explicit expressions for the currents one can check the following relations between different elements of
the conductance matrix  $-e
\partial \langle j_n(t) \rangle/\partial \mu_m$,  e.g. \cite{17,25}. Current conservation implies $\sum_n \partial \langle j_n(t)
\rangle/\partial \mu_m=0$, $n,m=L,R_1,R_2$. On the other hand, $\sum_m \partial \langle j_n(t) \rangle/\partial \mu_m=0$ only
for $\Delta_d-\Delta_L=\Delta_d-\Delta_{R_1}=\Delta_d-\Delta_{R_2}$. For other relations between the amplitudes $\Delta_d$ and
$\Delta_L$, $\Delta_{R_1}$, $\Delta_{R_2}$ we have:

\begin{eqnarray}
 \sum_k \partial \langle j_{R_1}(t) \rangle/\partial \mu_k=\sum_k \partial
\langle j_{R_2}(t) \rangle/\partial \mu_k=-{1\over 2}\sum_k \partial \langle j_{L}(t) \rangle/\partial \mu_k
\end{eqnarray}
for $\Delta_d-\Delta_{R_1}=\Delta_d-\Delta_{R_2}\neq\Delta_d-\Delta_{L}$, and

\begin{eqnarray}
\sum_k \partial \langle j_{R_1}(t) \rangle/\partial \mu_k\neq\sum_k \partial \langle j_{R_2}(t) \rangle/\partial \mu_k\neq
\sum_k \partial \langle j_{L}(t) \rangle/\partial \mu_k
\end{eqnarray}
for $\Delta_d-\Delta_{R_1}\neq\Delta_d-\Delta_{R_2}\neq\Delta_d-\Delta_{L}$.

\section{Results and discussion}

We consider the QD coupled with three, say the left, the first right and second right metal leads with the additional over-dot
(bridge) couplings between the left and right leads. We present the results for the time-averaged  currents and derivatives of
the average current with respect to the QD energy level in the presence of external microwave fields which are applied to the
dot and three leads, respectively. The time-dependent currents are also calculated in the case when the periodic
rectangular-pulse external fields are applied to each QD-lead barrier.

We have taken for $V_{LR}$ the values comparable with $V_{\vec k_\alpha d}$ and estimated $V_{\vec k_\alpha d}$ (assuming its
$\vec k$-independence, $V_{\vec k_\alpha d} \equiv V_\alpha = V$) using the relation $\Gamma_\alpha =
2\pi|V_\alpha|^2/D_\alpha$, where $D_\alpha$ is the $\alpha$-lead bandwidth and $D_\alpha = 100~\Gamma_\alpha$ ($\Gamma_L =
\Gamma_R = \Gamma$, $D_L = D_R = D$ was assumed). In our calculations we assumed $e=1$, all energies are given in $\Gamma$
units, time in $\hbar /\Gamma$ units, the current, its derivatives and frequency are given in $e\Gamma/\hbar, e^2\Gamma/\hbar$
and $\Gamma/\hbar$ units, respectively.

Firstly, we consider the time-averaged currents in the presence of the time-varying (harmonic case) external fields. Here we
give the explicit formula for the averaged current, $\langle j_L(t)\rangle$, performing the corresponding integrals in the
general formula given in Eq. (18), and compare it with the current $\langle j_L^{(2)}(t)\rangle$ flowing in the system of the QD
coupled with two leads only:
\begin{eqnarray}
\langle j_L(t)\rangle &=& \sum_{i=R_1,R_2} \left\{ {2x^2 \over (1+2x^2)^2}(\mu_L-\mu_i)  + {\Gamma \over 3\pi}{1-13x^2+4x^4
\over (1+2x^2)^3} \right. \nonumber\\ &\times & \sum_k \left[ J_k^2  \left( {\Delta_d-\Delta_L \over \omega} \right)
\arctan(h_L)- J_k^2 \left( {\Delta_d-\Delta_i \over \omega} \right) \arctan(h_i)  \right] \nonumber\\
&+& {\Gamma \over \pi}{x(1-2x^2) \over (1+2x^2)^3} \nonumber\\  &\times & \left. \sum_k \left[ J_k^2  \left( {\Delta_d-\Delta_L
\over \omega} \right) \ln(g_L)- J_k^2 \left( {\Delta_d-\Delta_i \over \omega} \right) \ln(g_i) \right] \right\},
\end{eqnarray}
where $h_i=\left( \mu_i-\varepsilon_d-\omega k+{2\Gamma x\over 1+2x^2} \right)/\left( {3\Gamma \over 2(1+2x^2)}\right)$,
$g_i=\left( \mu_i-\varepsilon_d-\omega k+{2\Gamma x\over 1+2x^2} \right)^2+\left( {3\Gamma \over 2(1+2x^2)}\right)^2$ and
$i=L,R_1, R_2$, whereas for $\langle j_L^{(2)}(t)\rangle$ we have:
\begin{eqnarray}
\langle j_L^{(2)}(t)\rangle &=& {2x^2 \over \pi(1+x^2)^2}(\mu_L-\mu_R)+ {\Gamma \over 2\pi}{1-6x^2+x^4 \over (1+2x^2)^3}
\nonumber\\ &\times & \sum_k \left[ J_k^2  \left( {\Delta_d-\Delta_L \over \omega} \right)
\arctan(h_L^{(2)})- J_k^2 \left( {\Delta_d-\Delta_R \over \omega} \right) \arctan(h_R^{(2)})  \right] \nonumber\\
&+& {\Gamma \over \pi}{x(1-x^2) \over (1+x^2)^3} \nonumber\\  &\times & \sum_k \left[ J_k^2  \left( {\Delta_d-\Delta_L \over
\omega} \right) \ln(g_L^{(2)})- J_k^2 \left( {\Delta_d-\Delta_i \over \omega} \right) \ln(g_R^{(2)}) \right],
\end{eqnarray}
where $h_i^{(2)}=\left( \mu_i-\varepsilon_d-\omega k+{\Gamma x\over 1+2x^2} \right)/\left( {\Gamma \over (1+2x^2)}\right)$ and
$g_i^{(2)}=\left( \mu_i-\varepsilon_d-\omega k+{\Gamma x\over 1+x^2} \right)^2+{\Gamma^2 \over (1+x^2)^2}$, $i=L,R$. Note, that
$\langle j_L(t)\rangle$ consists of the two terms and each term is similar in its structure to the current flowing in the QD-two
leads (QD-2L) system, $\langle j_L^{(2)}(t)\rangle$. However, due to the interference of the charge carriers propagating along
the different ways the arguments of the \emph{arctangens}  and \emph{logarytmic} functions are different and the individual
terms in $\langle j_L(t)\rangle$ and $\langle j_L^{(2)}(t)\rangle$ are weighted by different $x$-dependent factors.

\begin{figure}[tb]
\begin{center}
\includegraphics[width=0.6\columnwidth]{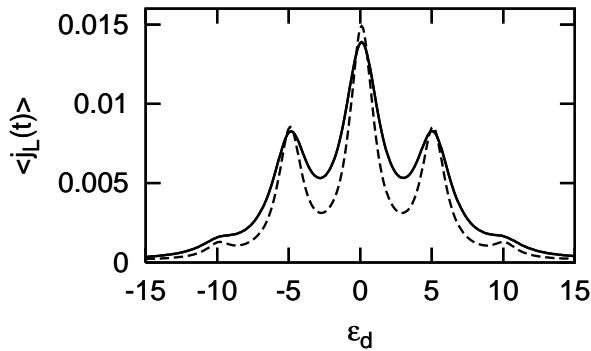}
\end{center}
\caption{The averaged current $\langle j_L(t)\rangle$ against $\varepsilon_d$ in the system of a QD coupled with three (solid
line) or two (broken line) leads. The microwave field is applied to the QD with $\Delta_d=6$ and $\omega=5$, $\mu_L=0.2$,
$\mu_{R_1}=\mu_{R_2}=0$, $\Gamma=1$, $V_{RL}=0$.} \label{fig2}
\end{figure}
In Figure \ref{fig2} we compare the averaged values of the current flowing from the left lead in the systems in which the QD is
coupled with three or two leads, the solid and broken curves, respectively. The external microwave field is applied only to the
QD and dc bias between the left and right leads is small in comparison with $\omega$, $\Delta_d$ and $\Gamma$. The coupling
$V_{LR}$ is assumed to be zero. In such a case the sidebands on the current curve are clearly visible. The width of the
corresponding peaks is smaller for the case of the QD coupled with two leads. Analyzing the expressions for $\langle
j_L(t)\rangle$ and $\langle j_L^{(2)}(t)\rangle$ one can obtain (for $\mu_L << \Gamma$ and $\mu_{R_1}=\mu_{R_2}=0$) the
subsequent peaks in $\langle j_L(t)\rangle$ (as functions of $\varepsilon_d$) in the form ${4 \over 9 \pi}J_k^2 \left( {\Delta_d
\over \omega}\right) \mu_L \left( 1+{\left(\varepsilon_d+\omega k \over 3\Gamma/2\right)^2 } \right)^{-1}$ with the FWHM equal
$3\Gamma$. For comparison, in the case of the QD coupled with two leads the corresponding peaks are described by the functions
${1 \over 2\pi}J_k^2 \left( {\Delta_d \over \omega}\right) \mu_L \left( 1+\left({\varepsilon_d+\omega k \over \Gamma}\right)^2
\right)^{-1}$ with the FWHM equal $2\Gamma$.

\begin{figure}[tb]
\begin{center}
\includegraphics[width=0.6\columnwidth]{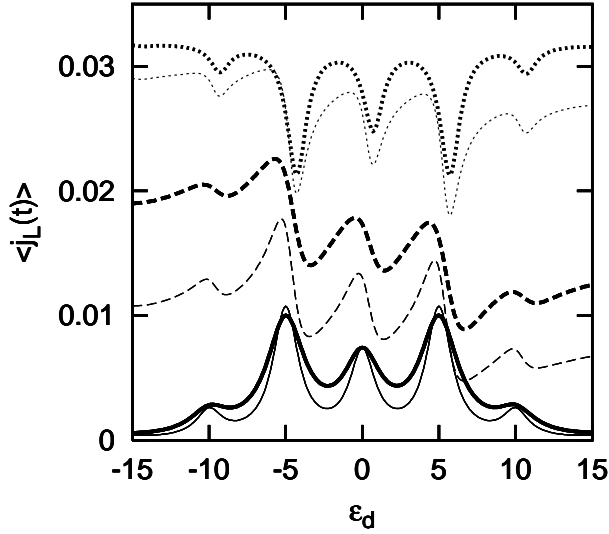}
\end{center}
\caption{The averaged current $\langle j_L(t)\rangle$ against $\varepsilon_d$ in the system of a QD coupled with three (thick
lines) or two (thin lines) leads. The solid, broken and dotted curves correspond to $x=0, 0.28$ and $0.7$, respectively.
$\Delta_d=8$, $\Delta_{R_1}=\Delta_{R_2}=\Delta_L=0$, $\mu_L=-\mu_{R_1}=-\mu_{R_2}=0.1$, $\omega=5$.} \label{fig3}
\end{figure}
Next, in Fig. \ref{fig3} we analyze the influence of the over-dot coupling between the left and right leads. Again, as in Fig.
\ref{fig2}, we consider an external microwave field applied only to the QD with $\Delta_d=8$ and $\mu_L=-\mu_{R_1}=-\mu_{R_2}$.
The thin (thick) lines correspond to the QD coupled with two (three) leads. We show $\langle j_L(t)\rangle$ for three values of
the inter-leads coupling strength represented by the parameter $x=\pi V_{LR}/D$. For $x=0$ we have well defined sideband peaks
as previously shown in Fig. \ref{fig2}. Next, we show the results for $x=\sqrt{13-\sqrt{153} \over 8} \approx 0.28$ and
$x=\sqrt{2}/2\approx 0.7$. Note, that the expression for $\langle j_L(t)\rangle$ (and $\langle j_L^{(2)}(t)\rangle$) consists of
three terms. The first term depends on the difference $\mu_L-\mu_i$ and does not depend on the QD energy level $\varepsilon_d$.
The second term corresponds to the Loretzian-type contribution to a given sideband (it disappears for $x\approx 0.28$) and the
last term corresponds to an asymmetric contribution (it disappears for $x\approx 0.7$), respectively. This last term influences
the sideband shape only for nonvanishing over-dot coupling between leads and is the most prominent sign of the interference
effects. For $x=\sqrt{2}/2$ this term disappears and the resulting sidebands have a form of symmetric dips due to the negative
value of the coefficient $\Gamma (1-13x^2+x^4) \over 3 \pi (1+2x^2)^3)$ in Eq. 27. In this case the nonresonant tunneling
channels modify the dips center position and its FWHM in comparison with the position and FWHM of sidebands presented for $x=0$.
Note, that for a QD coupled with two leads, the corresponding sidebands (the thin dotted line) are not fully symmetric curves as
in this case the last term of Eq. 28 disappears for $x=1$ and not for $x=0.7$ (c.f. ref. \cite{26}). For $x\approx 0.28$ the
corresponding sidebands are fully asymmetric curves as in this case the second term in Eq. 27  (which introduces asymmetry)
disappears. Again, for a QD coupled with two leads the corresponding sidebands (the thin broken curve) are described by not
fully asymmetric curves as the second term of Eq. 28, which gives a symmetric contribution to sidebands, disappears for
$x=\sqrt{3-2\sqrt{2}}\approx 0.38$ and not for $x =0.28$.

\begin{figure}[tb]
\begin{center}
\includegraphics[width=0.5\columnwidth]{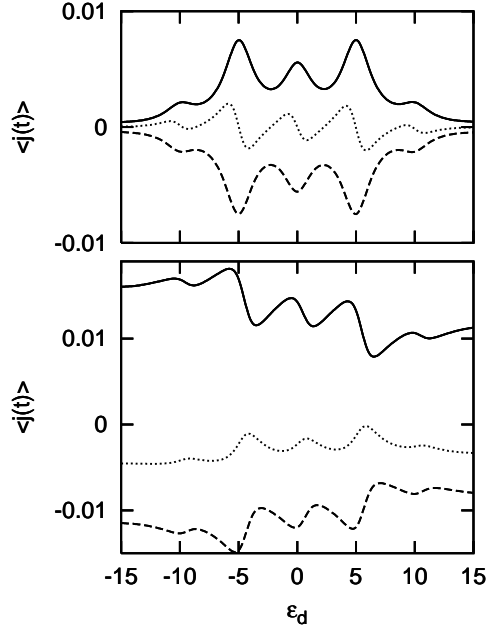}
\end{center}
\caption{The averaged currents $\langle j_L(t)\rangle$, $\langle j_{R_1}(t)\rangle$ and $\langle j_{R_2}(t)\rangle$ (solid,
broken and dotted lines, respectively) against $\varepsilon_d$ in the system of a QD coupled with three leads. The upper (lower)
panel corresponds to $x=0$ ($x=0.28$). $\Delta_d=8$ and other parameters as in Fig. \ref{fig3}. The values of $\langle
j_{R_2}(t)\rangle$ for $x=0$ (upper panel) have been multiplied by a factor of 20 for illustrating purposes.} \label{fig4}
\end{figure}

In the next Fig. \ref{fig4} we show the currents flowing in the QD-three leads (QD-3L) system for $x=0$ and $x=0.28$. For
vanishing values of the over-dot coupling the current $\langle j_L(t)\rangle$ is characterized by a sequence of the symmetric
peaks, but the current $\langle j_{R_2}(t)\rangle$ is a superposition of the asymmetric structures placed in the points where
the symmetric sidebands occur on $\langle j_L(t)\rangle$ curve. Analyzing the current $\langle j_{R_2}(t)\rangle$ according to
Eq. 29 we have for the parameters in Fig. \ref{fig4}:

\begin{eqnarray}
\langle j_{R_2}(t)\rangle &=& {\Gamma \over 3 \pi} \sum_{k} J_k^2 \left( {\Delta_d \over \omega} \right) \left[
\arctan(h_{R_2})- \arctan(h_L)  \right. \nonumber\\ &+& \left. \arctan(h_{R_2})- \arctan(h_{R_1})  \right],
\end{eqnarray}
where $h_i=\left( \mu_i-\varepsilon_d -\omega k \right)/{3\Gamma \over 2}$. One can see that each sideband is the sum of the
peak (two first terms in Eq. 29) and the dip (the last two terms in Eq. 29) resulting in the asymmetric structure shown in Fig.
\ref{fig4}.

\begin{figure}[tb]
\begin{center}
\includegraphics[width=0.4\columnwidth]{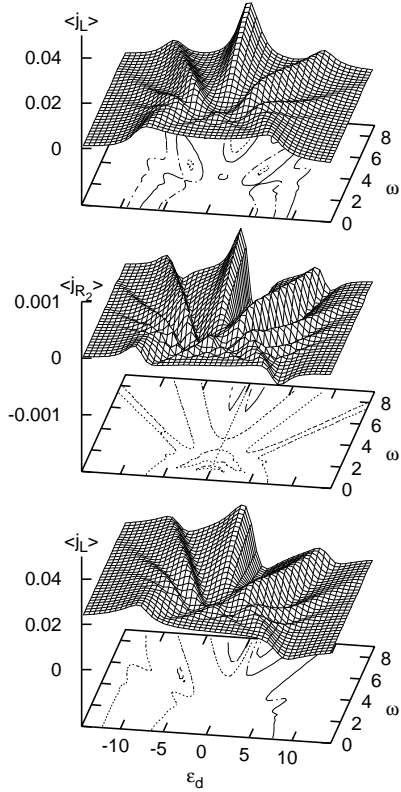}
\end{center}
\caption{The averaged currents against $\varepsilon_d$ and $\omega$. The upper (middle) panel shows $\langle j_L(t)\rangle$
($\langle j_{R_2}(t)\rangle$) for $x=0$ and the lower panel gives $\langle j_L(t)\rangle$ for $x=0.28$. $\Delta_d=8$,
$\Delta_{R_1}=\Delta_{R_2}=\Delta_L=0$, $\mu_L=-\mu_{R_1}=-\mu_{R_2}=0.2$.} \label{fig5}
\end{figure}
To learn more about the influence of the third electrode and additional over-dot coupling between leads we present in Fig.
\ref{fig5} the currents $\langle j_L(t)\rangle$ for $x=0$ and $x=0.28$ and $\langle j_{R_2}(t)\rangle$ for $x=0$. For vanishing
coupling between leads ($x=0$) the current $\langle j_L(t)\rangle$ exhibits a well known sideband structure for $\Gamma <
\omega$ and for small frequencies, $\omega < \Gamma$, the two broad maxima at $\varepsilon_d=\pm \Delta_d$ are present. At the
same time, the current $\langle j_{R_2}(t)\rangle$ exhibits the asymmetric structures centered  on the $(\varepsilon_d, \omega)$
plane at the points where photon sidebands occur on the $\langle j_L(t)\rangle$ curves. These asymmetric structures on the
$\langle j_{R_2}(t)\rangle$ curve exist also at $\omega < \Gamma$. On the other hand, similar structure of the $\langle
j_L(t)\rangle$ (the lower panel of Fig. \ref{fig5}) is obtained for $x=0.28$, i.e. we observe a number of asymmetric resonances
separated by the photon energy for $\omega > \Gamma$. Notice the similarity of both pictures, i.e. $\langle j_L(t)\rangle$
calculated for $x=0.28$ and $\langle j_{R_2}(t)\rangle$ for $x=0$, respectively. Note, however, the different scale for these
currents.

\begin{figure}[tb]
\begin{center}
\includegraphics[width=0.5\columnwidth]{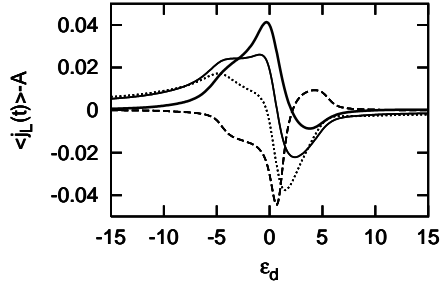}
\end{center}
\caption{The averaged current $\langle j_L(t)\rangle$ against $\varepsilon_d$ for different values of the over-dot coupling
strength between the left and right leads. The broken, dotted, thin solid and thick solid lines correspond to $x=0, 0.14,  0.28$
and $0.7$, respectively. $A=2x^2(2\mu_L-\mu_{R_1}-\mu_{R_2})/(1+2x^2)^2$ and $\Delta_{R_1}=3$,
$\Delta_{L}=\Delta_{R_2}=\Delta_d=0$, $\mu_L=-\mu_{R_1}=0.2$, $\mu_{R_2}=0$, $\omega=5$.} \label{fig6}
\end{figure}
In Fig. \ref{fig6} we present $\langle j_L(t)\rangle$ for different over-dot coupling assuming a strong asymmetry of the applied
microwave field (ac potential is applied only to the $R_1$-lead in the QD-3L system). The additional $R_2$-lead is characterized
by the chemical potential $\mu_{R_2}=(\mu_L+\mu_{R_1})/2=0$. For better demonstration of the influence of the ac potential on
$\langle j_L(t)\rangle$ we moved down each curve by the constant value $A=2x^2(2\mu_L-\mu_{R_1}-\mu_{R_2})/(1+2x^2)^2$, see Eq.
27. For $x=0$ we observe a shoulder on the left side of the main peak and a small negative current for the positive values of
$\varepsilon_d$. This picture is similar to the known results (for $x=0$) obtained experimentally and theoretically in the QD-2L
systems, e.g. Refs. \cite{3,26}. With the increasing over-dot coupling $V_{LR}$ the height of the main resonant peak decreases
and disappears at all for $x=0.28$ but the enhancement of the current for negative values of $\varepsilon_d$ (for
$\varepsilon_d\sim -5$) is more and more expressive. At the same time, for all values of $x$ we observe a negative current for
the small positive $\varepsilon_d$  with a greater absolute value for stronger coupling between leads. For greater values of $x$
the shape of the $\langle j_L(t)\rangle$ curve is changed dramatically and for $x=0.7$, $\langle j_L(t)\rangle$ becomes nearly
reversed in comparison with that calculated for $x=0$.

\begin{figure}[tb]
\begin{center}
\includegraphics[width=0.4\columnwidth]{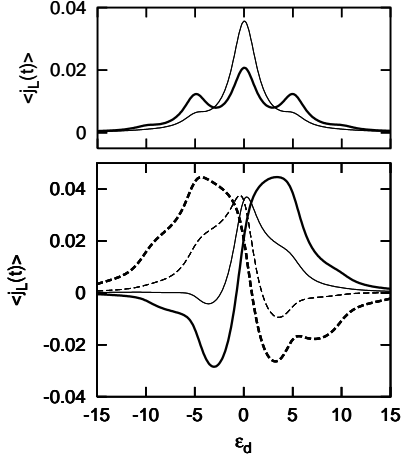}
\end{center}
\caption{The averaged current $\langle j_L(t)\rangle$ against $\varepsilon_d$ for the case of the microwave field applied only
to the QD with $\Delta_d=3$ (thin line) or $\Delta_d=6$ (thick line) - upper panel. The lower panel corresponds to the case of
the microwave field applied to the QD and $R_2$ lead with $\Delta_d=2\Delta_{R_2}$ and $\Delta_{R_2}=3$ or $6$ (thin or thick
lines, respectively). The broken lines show the results when the microwave field applied to the QD and $R_2$ lead are out of
phase (with the phase difference of $\pi$), $\mu_L=-\mu_{R_1}=0.2$, $\mu_{R_2}=0$, $\omega=5$.} \label{fig7}
\end{figure}
In Fig. \ref{fig7} we analyze the influence of the third lead (here named as $R_2$) on the current $\langle j_L(t)\rangle$ when
the external microwave field is applied to this lead and to the QD with $\Delta_d=2\Delta_{R_2}$. For comparison, we add in the
upper panel the results for $\langle j_L(t)\rangle$ obtained for the case when this additional third lead is not irradiated by
the microwave field. In this case, as before, see e.g. Fig. \ref{fig2}, we observe typical sidebands on the current curves (the
difference between the lead chemical potentials is small in comparison with the amplitude $\Delta_d$). However, after including
the third lead irradiated by the external microwave field the dependence of the current $\langle j_L(t)\rangle$ on the gate
voltage (or equivalently on the QD energy level position) is quite different - compare the thin or thick solid lines of the
upper and lower panels. For smaller values of $\Delta_{d}$ and $\Delta_{R_2}$, the averaged current $j_L$ is very similar to the
corresponding current $J_L$ obtained by applying the external microwave field only to one lead (see the broken line in Fig.
\ref{fig6}. These curves are, however, related between themselves by a relation $J_L(\varepsilon_d) \simeq j_L(-\varepsilon_d)$.
Now we can observe a small negative current at small negative values of $\varepsilon_d$ and some enhancement of the current on
the right side of the main peak. Similarly, the significant differences between the corresponding currents are observed also for
greater values of the amplitudes $\Delta_{d}$ and $\Delta_{R_2}$ (compare the thick solid lines in the upper and lower panels of
Fig. \ref{fig7}). Note, that very similar behavior of the current $\langle j_L(t)\rangle$ as the function of the gate voltage is
observed if we compare the case when the microwave field is applied only to one lead and the case when the microwave field is
applied simultaneously  to the QD and $R_2-$lead but with the phase difference of $\pi$ - compare the thin broken curve in Fig.
\ref{fig7} with the broken curve in Fig. \ref{fig6}.

\begin{figure}[tb]
\begin{center}
\includegraphics[width=0.4\columnwidth]{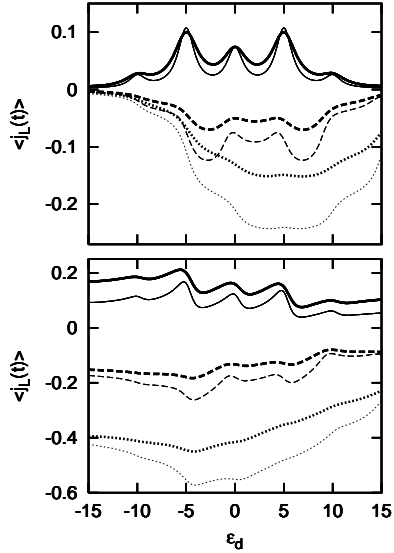}
\end{center}
\caption{The averaged current $\langle j_L(t)\rangle$ against $\varepsilon_d$ in the system of a QD coupled with three (thick
lines) or two (thin lines) leads. The upper (lower) panel corresponds to $x=0$ ($x=0.28$). In a QD-three leads system
 $\mu_L=0.1$, $\mu_{R_1}=-0.1$ and $\mu_{R_2}=-0.1, 4$ and $10$ (solid, broken and dotted curves, respectively).
 In a QD coupled with two leads  $\mu_L=0.1$, $\mu_{R}=-0.1, 4$ and $10$ (solid, broken and dotted curves, respectively). The
 values of solid curves have been multiplied by a factor of 10 for the illustrating purposes. $\omega=5$,
 $\Delta_d=8$, $\Delta_{R_1}=\Delta_{R_2}=\Delta_L=0$.} \label{fig8}
\end{figure}
Next, we consider the QD-2L system for different values of the chemical potential of the right lead ($\mu_{R_2}=-0.1, 4$ and 10)
and show the results of including the next lead ($R_1-$lead) to the system with $\mu_{R_1}=-0.1$. In Fig. \ref{fig8} we present
the results obtained for $\langle j_L(t)\rangle$ in the case when the microwave field is applied only to the QD. For the case
$\mu_L=-\mu_{R_1}=-\mu_{R_2}=0.1$ and $x=0$ the sidebands are very clearly visible and the corresponding peaks are lower and
broader for the QD-3L system as we discussed before (the upper panel). For the non-zero over-dot coupling between the left and
right leads the sideband peaks get asymmetric forms and for $x=0.28$ they become fully asymmetric (the lower panel). The current
for the QD-2L system is also composed of a number of asymmetric components although now these forms are fully asymmetric for
$x=0.38$ as we know from the earlier discussion. For greater values of $\mu_{R_2}$, the corresponding current flowing in the
QD-2L system achieves (for vanishing, as well as for nonzero coupling between leads) greater negative values and its dependence
on the QD energy level position is well-marked in comparison to the results characterizing the QD-2L system.

\begin{figure}[tb]
\begin{center}
\includegraphics[width=0.4\columnwidth]{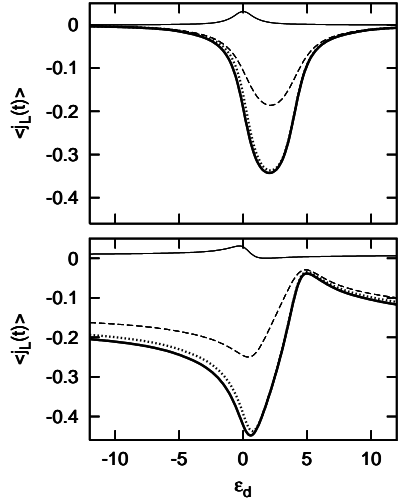}
\end{center}
\caption{The averaged current $\langle j_L(t)\rangle$ against $\varepsilon_d$ for the QD-2L systems: $\mu_L=-\mu_R=0.1$ - thin
solid lines and $\mu_L=0.1$, $\mu_R=4$ - thick solid lines and the QD-3L system: $\mu_L=-\mu_{R_1}=0.1$, $\mu_{R_2}=4$ - broken
lines. The upper (lower) part corresponds to $x=0$ ($x=0.28$) and  $\omega=5$, $\Delta_d=1$,
$\Delta_{R_1}=\Delta_{R_2}=\Delta_L=0$. The dotted lines correspond to the sum of the currents flowing in two QD-2L systems.}
\label{fig9}
\end{figure}
In order to emphasize the influence of the additional lead on the currents we show in Fig. \ref{fig9} the current $\langle
j_L(t)\rangle$ for QD-2L and QD-3L systems calculated for the parameters for which the corresponding curves are relatively
simple. We assumed the small amplitude of the QD energy level oscillations, $\Delta_d=1$, and $\omega=5$, for which (for $x=0$)
only the central peak corresponding to elastic tunneling is visible on the $\langle j_L(t)\rangle$ curves. We show $\langle
j_L(t)\rangle$ obtained for two different QD-2L systems which can be viewed as components of more complicated QD-3L system. We
observe that due to the interference effects, the current $\langle j_L(t)\rangle$ flowing in the QD-3L system is not simply a
sum of currents (dotted lines) flowing in the corresponding QD-2L systems. The difference between this sum and the current
corresponding to the QD-3L system is relatively large and exists independently of the coupling between leads.

\begin{figure}[tb]
\begin{center}
\includegraphics[width=0.31\columnwidth]{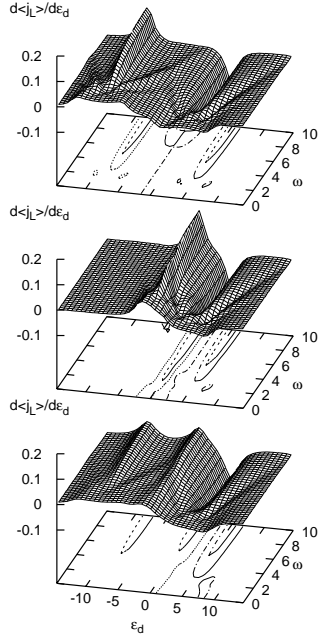}
\end{center}
\caption{The averaged current derivatives $d\langle j_L(t)\rangle/d\varepsilon_d$ against $\varepsilon_d$ and $\omega$ for
$x=0$. We compare the results obtained for the QD coupled with two leads - the upper (middle) panel - for $\mu_L=5$,
$\mu_{R}=-8$, $\Delta_L=8$, $\Delta_{d}=4$, $\Delta_R=0$ ($\mu_L=5$, $\mu_{R}=0$, $\Delta_L=8$, $\Delta_{d}=4$, $\Delta_R=2$)
with the results obtained for the QD coupled with three leads - the lower panel - $\Delta_L=8$, $\Delta_d=4$, $\Delta_{R_1}=2$,
$\Delta_{R_2}=0$, $\mu_L=5$, $\mu_{R_1}=0$, $\mu_{R_2}=-8$, $\omega=5$.} \label{fig10}
\end{figure}
In order to present more information about the differences between the electron transport in the QD-3L and QD-2L systems we
display in Fig. \ref{fig10} the derivative $d\langle j_L(t)\rangle/d\varepsilon_d$ as a function of $\varepsilon_d$ and
$\omega$. The lowest panel corresponds to the QD-3L system and two other panels correspond to the QD-2L systems. These QD-2L
systems are characterized by such parameters that combined together give us the considered QD-3L system. One can see, that the
considered characteristics of the electron transport in the QD-3L system are not simply the algebraic sum of the corresponding
curves of both QD-2L systems. In all three cases shown in Fig \ref{fig10}, the position of the corresponding minima and maxima
(along the $\varepsilon_d-$axis) can be identified with the values of the leads chemical potentials. However, the corresponding
structures are less clear in the case of the QD-3L system in comparison with those for the QD-2L models.

\begin{figure}[tb]
\begin{center}
\includegraphics[width=0.5\columnwidth]{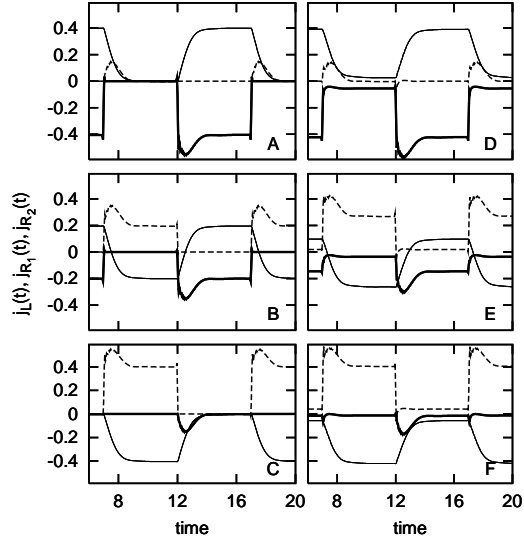}
\end{center}
\caption{The time-dependent current flowing in the system of a QD coupled with the three leads: $L$, $R_1$ and $R_2$. The
$L$-lead is coupled with the QD only - the left panels and with the QD and two other leads, $V_{LR_1}=V_{LR_2}=4$ - the right
panels. The couplings between the QD and $R_1, R_2$ leads are changed periodically. The upper, middle and lower panels
correspond to $\mu_L=3,0$ and -3, respectively. $-\mu_{R_1}=\mu_{R_2}=3, \varepsilon_d=0$. The thin, thick and broken curves
correspond to $j_L, j_{R_1}$ and $j_{R_2}$ currents, respectively.} \label{fig11}
\end{figure}
In the next step of our investigations of the electron transport in the QD-3L systems we consider the time-dependent currents
flowing in response to the time-dependent barriers between the QD and leads or in response to suddenly removed (or inserted)
connection of the QD with one of the leads. As a first we assume a rectangular-pulse modulation applied to the QD-$R_1$ lead and
QD-$R_2$ lead barriers. We assume that these modulations are with a phase difference of $\pi$. In the first (second) half-cycle,
$V_{dR_1}=0$ ($V_{dR_2}=0$) and the QD is coupled only to the $R_2$ lead ($R_1$ lead). In addition, the QD is coupled to the
next, say $L-$lead, with a constant value $V_{dL}$. In the following we consider the time-dependent currents $ j_L(t)$, $
j_{R_1}(t)$ and $j_{R_2}(t)$ for the three specific conditions: $\mu_L=\mu_{R_2}$, $\mu_L=(\mu_{R_1}+\mu_{R_2})/2$ and
$\mu_L=\mu_{R_1}$. In addition, we assume $\mu_{R_2}=-\mu_{R1}=3$, $\varepsilon_d=\mu_{R_1}$ and take for the period of the
considered barrier modulation $T=5$. In a such case we integrate numerically the corresponding set of the differential equations
for the matrix elements of the evolution operator and in the next step calculate the currents according to the formula
$j_a(t)=-edn_a(t)/dt$, where $a=L,R_1$ or $R_2$ and $n_a(t)$ is given in  Eq. 5 (or similar to it). We checked that the QD
charge hardly depends on the additional $V_{LR}$ couplings. Although the QD charge is almost insensitive to the additional
over-dot couplings the currents demonstrate such dependence. Especially visible are the differences for the case when the
chemical potential $\mu_L$ of the third electrode $L$ lies between the  chemical potentials of two other leads, see Fig.
\ref{fig11} B,E. For other values of $\mu_L$, the influence of the over-dot tunneling channels for the parameters considered
here is smaller. Note, that after abrupt changing of the coupling between the QD and $R_1$, or $R_2$ leads the currents $j_L,
j_{R_1}$ and $j_{R_2}$ rapidly change too, and after a short time reach the steady values. The QD coupled with three leads could
be considered as the three-state system. We observe that for some values of the lead chemical potentials the currents change
their values periodically e.g. from  zero to the positive value (see $j_{R_2}$ in Fig. \ref{fig11}B), from zero to the negative
value (see $j_{R_1}$ in Fig. \ref{fig11}B) or from the negative to the positive value (see $j_L$ in Fig. \ref{fig11}B).The
additional couplings between leads modify the values of the currents but the qualitative picture remains the same. Note, that in
the first moment after abrupt changing of the coupling between the QD and $R_1$ or $R_2$ lead we have
$j_L(t)+j_{R_1}(t)+j_{R_2}(t)\neq 0$ as in this case $dn_d(t)/dt \neq 0$ (not shown here). After some delay time the QD charge
stabilizes around its equilibrium value, the currents tend to constant values and their sum is equal to zero.

\begin{figure}[tb]
\begin{center}
\includegraphics[width=0.5\columnwidth]{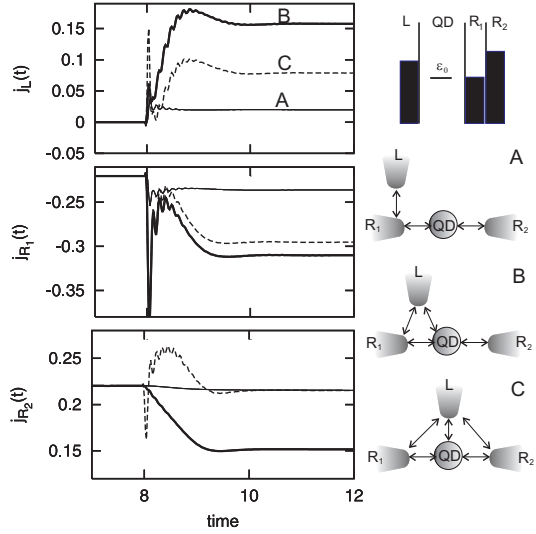}
\end{center}
\caption{The time-dependent currents flowing from the $L$, $R_1$ and $R_2$ leads in the system shown in Fig. \ref{fig1}. The
thin solid, thick solid and broken curves correspond to different couplings of the $L$ lead to the other elements of the system:
($V_{LR_1}=4$, $V_{Ld}=V_{LR_2}=0$), ($V_{LR_1}=4$, $V_{Ld}=4$, $V_{LR_2}=0$) and ($V_{LR_1}=V_{Ld}=V_{LR_2}=4$), respectively.}
\label{fig12}
\end{figure}
In the last step we consider the response of the currents to the abrupt inclusion into the QD-2L system of the third electrode
(in our case, $L$-lead). The results are presented in Fig. \ref{fig12} together with the schematic view of the subsequent tunnel
connections between the QD and the three leads. We show the time-dependent currents $j_L(t)$, $j_{R_1}(t)$ and $j_{R_2}(t)$
corresponding to the three different ways of inclusion of the $L$-lead. We assumed the chemical potentials
$\mu_{R_1}=-\mu_{R_2}$, $\mu_L=(\mu_{R_1}+\mu_{R_2})/2$ and $\varepsilon_d=\mu_{R_2}$. Consider the current $j_{R_2}$ flowing
from the $R_2$-lead characterized by the highest chemical potential $\mu_{R_2}$. Before adding to the system of the $L$-lead the
current $j_{R_2}$ has a constant value and flows from $R_2$-lead through the QD energy level to $R_1$-lead. When the $L$-lead is
included into the system (the tunneling coupling $V_{LR_1}$ changes abruptly at $t=8$ from zero to nonzero value) the current
$j_{R_2}$ is almost unchanged - its value decreases slightly without  any transients at short times after the time $t=8$. Next,
we consider the case when the $L$-lead is abruptly connected simultaneously with the QD and $R_1$-lead. Now the current
$j_{R_2}$ decreases significantly during the short time after the moment of inclusion of $L$-lead and settles to its constant
value. Note, that $j_{R_2}$ decreases despite the additional charge transfer channel between $R_2$ and $L$ leads (through the QD
energy level). The destructive interference appears in this case as we have two transmission channels for tunneling electrons
between $R_2$ and $R_1$ leads. However, the constructive interference is visible if we consider the next case when the $L$-lead
is abruptly coupled with the QD-2L system assuming nonzero values of $V_{LR_1}$, $V_{LR_2}$ and $V_{Ld}$. Now, we have one
additional charge transfer channel (from $R_2$-lead) in comparison with the former case. The current $j_{R_2}$ rapidly increases
with some fluctuations and after the time $\sim \hbar/\Gamma$ decreases to the constant value greater than in QD-2L system.
Similar analysis can be made considering the currents $j_{L}$ and $j_{R_1}$ although the transient current changes are more
visible now at short times after the abrupt inclusion of additional electron tunneling channels. The above discussion concerns
the specific values of the lead chemical potentials and the position of the QD energy level. Nevertheless, the similar
qualitative conclusions can be made also for other values characterizing the  considered system.

\section{Conclusions}

We have studied the time-dependent tunneling transport through the QD coupled with three metal leads using the evolution
operator technique. The time-dependent QD charge and currents were determined in terms of the appropriate evolution operator
matrix elements. Applying the wide band limit to the integrodifferential equations satisfied by the evolution operator matrix
elements we were able to give the analytical expressions for the time-averaged currents and differential conductance. We
considered the external harmonic microwave fields applied to different parts of the considered system, as well as, the
rectangular-pulse modulation imposed on different QD-leads barriers. In addition, we have studied also the time dependence of
the currents due to abrupt inclusion into the QD-two leads system of the third electrode. We have considered also the effect of
the additional couplings between leads (we coupled one of the leads with the other two leads) on the conductance and current
flowing in
the system. Our  main results can be summarized as follows:\\
- For the vanishing nonresonant tunneling path, $V_{LR}=0$, and for the parameters for which the photon-assisted sidebands are
clearly visible on the $\langle j_L(t)\rangle$ curve the subsequent sideband peaks have the Lorentz-type form with the FWHM
equal $3\Gamma$ in comparison with $2\Gamma$ for the QD-two leads system. For the increasing value of $V_{LR}$ the form of the
sidebands transforms from the Lorentz-type to the fully asymmetric form for $x=\sqrt{(13-\sqrt{153})/8} \simeq 0.28$. For
greater $x$ the form of the sidebands changes and gains again the Lorentz-type shape for $x=\sqrt{2}/2$. For the QD coupled with
two leads the corresponding values of $x$ are equal to $\sqrt{8-\sqrt{52}/3} \simeq 0.38$ and $1$, respectively. \\
- For the vanishing $V_{LR}$ the differential conductance curve, e.g. $d\langle j_L(t)\rangle/d\mu_L$, possesses the sidebands
of the Lorentz-type localized at $\mu_L=\varepsilon_d\pm \omega k$. For $V_{LR}\neq 0$ these sidebands are described by the
superposition of two parts, the Lorentz-type and the asymmetric one centered at $\varepsilon_d=\mu_L\pm \omega k + 2 \Gamma x/
(1+2x^2)$ weighed by the factors ${\Gamma (1-13x^2+4x^4) \over \pi (1+2x^2)^4}J_k^2 \left( {\Delta_d-\Delta_L \over \omega}
\right)$ and ${4\Gamma x (1-2x^2) \over (1+2x^2)^3}J_k^2 \left( {\Delta_d-\Delta_L \over \omega} \right)$, respectively. For the
QD-two leads system the corresponding $k-$th sidebands are centered at $\varepsilon_d=\mu_L\pm \omega k + \Gamma x/ (1+x^2)$ and
their symmetric and asymmetric parts are weighted by the factors ${\Gamma^2 (1-6x^2+x^4) \over \pi (1+x^2)^4}J_k^2 \left(
{\Delta_d-\Delta_L \over \omega} \right)$ and ${2\Gamma x(1-x^2) \over \pi (1+x^2)^3}J_k^2 \left( {\Delta_d-\Delta_L \over
\omega} \right)$, respectively.\\
- The symmetry properties of the sidebands corresponding  to the current flowing from the given lead depend on the position of
the chemical potential of this lead in comparison with the chemical potentials of the other two leads. Taking, for example
$V_{LR}=0$ and $\mu_{R_2}$ localized in the middle between $\mu_L$ and $\mu_{R_1}$ one can observe on the $\langle
j_L(t)\rangle$ curve the sidebands of nearly regular (Lorentz-type) forms while the sidebands on the $\langle j_{R_2}(t)\rangle$
curve have asymmetric structures. However, in the presence of the nonresonant tunneling path the sidebands on the  $\langle
j_L(t)\rangle$ curve change their form and for $x\approx 0.28$ they have a fully asymmetric shape. On the other hand, the
sidebands on the $\langle j_{R_2}(t)\rangle$ curve have for $x\approx 0.28$ a nearly Lorentz-like shape. \\
- Especially large interference effects can be observed if we compare the current flowing in the QD-3L system with the sum of
currents flowing in the two QD-2L systems which can be viewed as the components of the considered more complicated QD-3L system.
The difference between them is relatively large independently of the over-dot coupling between leads (see Fig. \ref{fig9})\\
- In the case of strong asymmetry of the applied external field ($\Delta_{R_1}\neq0, \Delta_L=\Delta_d=\Delta_{R_2}=0$) we
observe for $x=0$ a shoulder on the left side of the main resonant peak on the $\langle j_L(t)\rangle$ curve vs. the gate
voltage. With the increasing over-dot coupling between $L$ and $R_1, R_2-$leads the main resonant peak disappears and transforms
in a dip for strong coupling $V_{LR}$. At the same time a shoulder of this curve increases with the increasing $V_{LR}$.\\
- Let us consider the time dependence of currents flowing in response to the time-dependent barriers between the QD and two
leads (we assume a constant coupling of the QD with the third lead). For the assumed  rectangular-pulse modulation applied to
the QD-$R_1$ lead and QD-$R_2$ lead barriers one can consider the QD-3L system as a three-state one. For example, the currents $
j_L(t)$, $j_{R_1}(t)$ and $j_{R_2}(t)$ change their values periodically between zero and positive, positive and zero and
negative and positive values, respectively (see Fig. 11B). The additional couplings between the leads introduce only small
quantitative changes.
\\

{\bf Acknowledgements:} The work of one of us (RT) has been partially supported by the KBN grant No. PBZ-MIN-008/P03/2003.

\newpage
\centerline{\large\bf Appendix}

\bigskip
In this section we present the derivations of the general equations satisfied by the required functions $U_{dd}(t,t_0)$, $U_{d
\vec k} (t,t_0)$, $U_{d,\vec r/\vec q} (t,t_0)$ and $U_{\vec k d} (t,t_0)$, $U_{\vec k_1,\vec k_2} (t,t_0)$, $U_{\vec k,\vec
r/\vec q} (t,t_0)$ needed for the calculations of the QD charge $n_d(t)$ and the currents flowing in the considered system. In
the next step, using the WBL approximation we simplify these equations and give the analytical solutions for them. Let us begin
from the derivation of the integro-differential equation satisfied by $U_{dd}(t,t_0)$. Writing down the formal solution of Eq.
(11)
\begin{eqnarray}
U_{\vec q/\vec r d}(t,t_0) &=& -i\int\limits^t_{t_0}dt_1 \tilde V_{\vec q/\vec r d}
 (t_1) U_{dd}(t_1,t_0)  \nonumber \\
&-& i\int\limits^t_{t_0} dt_1 \sum_{\vec k} \tilde V_{\vec q/\vec r\vec k} (t_1)
 U_{\vec k d} (t_1,t_0)
\end{eqnarray}
and inserting them to the formal solution of Eq. (10)
\begin{equation}
U_{\vec k d}(t,t_0) = -i\int\limits^t_{t_0} dt_1 \tilde V_{\vec k d}(t_1) U_{dd} (t_1,t_0)
 - i\int\limits^t_{t_0} dt_1 \sum_{\vec p = \vec q,\vec r} \tilde V_{\vec k\vec p}
 (t_1) U_{\vec p d} (t_1,t_0)\,,
\end{equation}
one can obtain after straightforward calculations the function $U_{\vec kd}(t,t_0)$ expressed in the terms of $U_{dd}(t,t_0)$,
only
\begin{eqnarray}
 U_{\vec kd} (t,t_0) &=& (-i) \int\limits^t_{t_0}dt_1 \tilde V_{\vec kd}(t_1)
 U_{dd}(t_1,t_0) + (-i)^2 \int\limits^t_{t_0}\int\limits^{t_1}_{t_0} dt_1 dt_2\,
 R_{\vec kd} (t_1,t_2) U_{dd}(t_2,t_0)  \nonumber\\
&+& \sum^\infty_{j=2} (-i)^{2j} \int\limits^t_{t_0}\int\limits^{t_1}_{t_0}
\ldots \int\limits^{t_{2j-1}}_{t_0}dt_1 \ldots dt_{2j} \sum_{\vec k_1,\vec k_2,
\ldots,\vec k_{j-1}} R_{\vec k\vec k_1}(t_1,t_2)\, R_{\vec k_2
\vec k_3}(t_3,t_4)\ldots \nonumber \\
&& \hspace{3cm} R_{\vec k_{j-1}d} (t_{j-1},t_j) U_{dd}(t_j,t_0) \nonumber\\
&+& \sum^\infty_{j=1}(-i)^{2j+1} \int\limits^t_{t_0}\int\limits^{t_1}_{t_0}
\ldots \int\limits^{t_{2j}}_{t_0} dt_1 \ldots dt_{2j+1} \sum_{\vec k_1,\vec k_2,
\ldots,\vec k_j} R_{\vec k\vec k_1}(t_1,t_2)\, R_{\vec k_2
\vec k_3}(t_3,t_4)\ldots  \nonumber \\
&& \hspace{3cm} \tilde V_{\vec k_jd}(t_{2j+1}) U_{dd} (t_{2j+1},t_0)\,,
\end{eqnarray}
where
\begin{equation}
 R_{ij}(t_1,t_2) = \sum_{\vec  q} \tilde V_{i\vec q}(t_1) \tilde V_{\vec q j}
(t_2) + \sum_{\vec r} \tilde V_{i\vec r}(t_1) \tilde V_{\vec r j}(t_2)
\end{equation}
We remember, that the wave vectors $\vec k, \vec q$ and $\vec r$ correspond to the left lead and the first and second right
leads, respectively. Inserting into Eq. 9 the expressions for the functions $U_{\vec q d}(t,t_0)$, $U_{\vec r d}(t,t_0)$ and
$U_{\vec k d}(t,t_0)$, Eqs. (17, 18), we can write the integro-differential equation for $U_{dd}(t,t_0)$ in the form
\begin{eqnarray}
{\partial U_{dd}(t,t_0)\over\partial t} &=& -\int\limits^t_{t_0} dt_1
 \left[\sum_{\vec k}\tilde V_{d\vec k}(t) \tilde V_{\vec kd}(t_1) + R_{dd}(t,t_0)
\right] U_{dd}(t_1,t_0) \nonumber\\
&+& \sum^\infty_{j=1}(-i)^{2j-1} \int\limits^t_{t_0}\int\limits^{t_1}_{t_0}
 \ldots \int\limits^{t_{2j-1}}_{t_0} dt_1 \ldots dt_{2j} \sum_{\vec k_1,\vec k_2,
\ldots,\vec k_j} [R_{d\vec k_1}(t,t_1) R_{\vec k_1\vec k_2} (t_2,t_3)
 \ldots\nonumber\\
&& R_{\vec k_{j-1}\vec k_j}(t_{2j-2},t_{2j-1}) \tilde V_{\vec k_jd}(t_{2j}) \nonumber\\ &+&
 \tilde V_{d\vec k_1}(t) R_{\vec k_1\vec k_2}(t_1,t_2)\ldots R_{\vec k_jd}
(t_{2j-1},t_{2j})] U_{dd}(t_{2j},t_0) \nonumber\\
&+& \sum^\infty_{j=1}(-i)^{2j} \int\limits^t_{t_0}\int\limits^{t_1}_{t_0}
 \ldots \int\limits^{t_{2j}}_{t_0} dt_1\ldots dt_{2j+1} \sum_{\vec k_1\vec k_2
 \ldots \vec k_j} [R_{d\vec k_1}(t,t_1) R_{\vec k_1\vec k_2} (t_2,t_3) \ldots \nonumber\\
&& R_{\vec k_{j-1}\vec k_j} (t_{2j-2},t_{2j-1}) \, R_{\vec k_jd}(t_{2j},t_{2j+1}) \\
&+& \sum_{\vec k_{j + 1}} \tilde V_{d\vec k_1}(t) \, R_{\vec k_1\vec k_2}(t_1,t_2) \ldots R_{\vec k_j,\vec
k_{j+1}}(t_{2j-1},t_{2j}) \tilde V_{\vec k_{j+1}d} (t_{2j+1})] U_{dd}(t_{2j+1},t_0) \,. \nonumber
\end{eqnarray}
This rather untractable general equation can be greatly simplified using the WBL approximation. Assuming $V_{d\vec r} = V_{d\vec
q} = V_{d\vec k} \equiv V$, $V_{\vec r\vec k} = V_{\vec q\vec k} \equiv V_{RL}$, the multidimensional time integrations and
summations over the wave vectors can be performed giving in result
\begin{equation}
  {\partial U_{dd}(t,t_0)\over \partial t} = -C_1\, U_{dd}(t,t_0) \,,
\end{equation}
with the solution
\begin{equation}
 U_{dd}(t,t_0) = {\rm exp} (-C_1(t - t_0))\,.
\end{equation}
Here $ C_1 = {3\over 2}\Gamma - \Gamma {3x^2 + 2ix \over 1 + 2x^2}$ and $x = \pi V_{RL}/D$, $D$ being the bandwidth of the lead
energy band ($D_{R_1}=D_{R_2}=D_{L}=D$). The function $U_{\vec kd}(t,t_0)$ given in Eq. (19) can be reduced within the WBL
approximation to the form
\begin{equation}
 U_{\vec kd}(t,t_0) = -C_2 \int\limits^t_{t_0} dt_1 \tilde V_{\vec kd}(t_1)
 U_{dd}(t_1,t_0)\,,
\end{equation}
where
\begin{equation}
C_2 = {i + 2x\over 1 + 2x^2}\,.
\end{equation}
In order to calculate $U_{d\vec k}(t,t_0)$ we write down accordingly with Eq. (3) the corresponding set of coupled differential
equations for the functions $U_{d \vec k}(t,t_0)$, $U_{\vec k_1\vec k_2}(t,t_0)$, $U_{\vec q\vec k}(t,t_0)$ and $U_{\vec r\vec
k}(t,t_0)$. The subsequent steps of the calculations are similar to those performed in the derivation of Eq. (34). Inserting the
formal solutions for $U_{\vec k_1\vec k_2}(t,t_0)$, $U_{\vec q/\vec r,\vec k}(t,t_0)$
\begin{eqnarray}
&& U_{\vec k_1\vec k_2}(t,t_0) = \delta_{\vec k_1\vec k_2} - i\int\limits^t_{t_0}
 dt_1 \left[\tilde V_{\vec k_1d}(t_1) U_{d\vec k_2} (t_1,t_0) +
  \sum_{\vec p = \vec q\vec r} \tilde V_{\vec k_1\vec p}(t_1) U_{\vec p\vec k_2}
 (t_1,t_0)\right] \,, \nonumber\\ \\
&& U_{\vec p\vec k}(t,t_0) =  - i\int\limits^t_{t_0}
 dt_1 \left[\tilde V_{\vec pd}(t_1) U_{d\vec k} (t_1,t_0) +
  \sum_{\vec k_1} \tilde V_{\vec pk_1}(t_1) U_{\vec k_1\vec k} (t_1,t_0)\right] \,,
\end{eqnarray}
into the differential equation satisfied by $U_{d\vec k}(t,t_0)$ one obtains the derivative $\partial U_{d\vec
k}(t,t_0)/\partial t$ expressed in terms of $U_{d\vec k}(t,t_0)$ and $U_{\vec k_1\vec k_2}(t,t_0)$. On the other hand, on the
basis of Eqs. (33, 34) the function $U_{\vec k_1\vec k_2}(t,t_0)$ can be represented in the form containing only $U_{d\vec
k}(t,t_0)$. Finally, one obtains
\begin{eqnarray}
 {\partial U_{d\vec k}(t,t_0)\over\partial t} &=& -i\tilde V_{d\vec k}(t) +
 (-i)^2\int\limits^t_{t_0} dt_1 \left[\sum_{\vec q_1} \tilde V_{d\vec q_1}
 \tilde V_{\vec q_1\vec k}
 (t_1) + \sum_{\vec r_1} \tilde V_{d\vec r_1}(t) \tilde V_{\vec r_1\vec k}(t_1)
\right] \nonumber\\
&+& (-i)^3 \int\limits^t_{t_0} \int\limits^{t_1}_{t_0} dt_1 dt_2 \left[\sum_{\vec k_1,\vec q_1}
 \tilde V_{d\vec k_1}(t) \tilde V_{\vec k_1\vec q_1}(t_1) \tilde V_{\vec q_1\vec k}
 (t_2) + \sum_{\vec k_1,\vec r_1} \ldots  \right] \nonumber \\
&+& (-i)^4 \int\limits^t_{t_0} \int\limits^{t_1}_{t_0} \int\limits^{t_2}_{t_0}
 dt_1 dt_2 dt_3 \left[\sum_{\vec q_1,\vec k_1,\vec q_2}
 \tilde V_{d\vec q_1}(t) \tilde V_{\vec q_1\vec k_1}(t_1) \tilde V_{\vec k_1\vec q_2}
 (t_2) \tilde V_{\vec q_2\vec k}(t_3) \right. + \nonumber\\
&& \hspace{3cm}\left. \sum_{\vec r_1,\vec k_1\vec q_2} \ldots + \sum_{\vec q_1\vec k_1
\vec r_2} \ldots + \sum_{\vec r_1\vec k_1\vec r_2} \ldots \right] + \ldots \nonumber\\
&+& (-i)^2 \int\limits^t_{t_0} dt_1 \left[\sum_{\vec k_1} \tilde V_{\vec d\vec k_1}(t)
 \tilde V_{\vec k_1d} (t_1)  + \sum_{\vec q_1} \tilde V_{d\vec q_1}(t)
  \tilde V_{\vec q_1d}(t_1)\right.  \nonumber\\
&&\left. \hspace{3cm} +\sum_{\vec r_1}\tilde V_{d\vec r_1}(t) \tilde V_{\vec r_1d}(t_1)\right]
  U_{d\vec k}(t_1,t_0)  \nonumber \\
&+& (-i)^3 \int\limits^t_{t_0} \int\limits^{t_1}_{t_0} dt_1 dt_2
  \left[\sum_{\vec k_1,\vec q_1} \tilde V_{d\vec k_1}(t) \tilde V_{\vec k_1\vec q_1}(t_1) \tilde V_{\vec q_1d}(t_2) \right.\nonumber\\
&&\hspace{3cm}\left. +\sum_{\vec k_1,\vec r_1} \ldots + \sum_{\vec q_1\vec k_1}
   \ldots + \sum_{\vec r_1\vec k_1} \ldots \right] U_{d\vec k}(t_2,t_0) + \ldots
\end{eqnarray}
In the WBL approximation this equation reduces to the form
\begin{equation}
 {\partial U_{d\vec k}(t,t_0)\over\partial t} = -C_2\tilde V_{d\vec k}(t) + C_3 U_{d\vec k}
 (t,t_0)
\end{equation}
with the solution
\begin{equation}
 U_{dk}(t,t_0) = -C_2 \int\limits^t_{t_0} dt_1 \tilde V_{d\vec k}(t_1) \,,
 {\rm ep} (-C_3(t - t_1))
\end{equation}
where $C_3 = \Gamma (4ix - 3)/2(1 + 2x^2)$. Taking into account Eqs. (33, 34) the function $U_{\vec k_1\vec k_2}(t,t_0)$ can be
written in terms of $U_{d \vec k_2}(t,t_0)$
\begin{eqnarray}
 U_{\vec k_1\vec k_2}(t,t_0) &=& \delta_{\vec k_1\vec k_2} - C_2 \int\limits^t_{t_0}
 dt_1 \tilde V_{\vec k_1 d} (t_1) U_{d\vec k_2}(t_1,t_0)  \nonumber \\
&&  - {2V_{LR} \cdot x\over 1 + 2x^2}\int\limits^t_{t_0} dt_1 \, e^{i(\varepsilon_{\vec k_1} -
 \varepsilon_{\vec k_2})(t_1 - t_0)} \,.
\end{eqnarray}

In order to calculate $U_{d \vec q}(t,t_0)$ we first write down the coupled set of equations (on the basis of Eq. (3)) satisfied
by the functions $U_{d\vec q}(t,t_0)$,
 $U_{\vec q_1\vec q_2}(t,t_0)$,  $U_{\vec r\vec q}(t,t_0)$
and $U_{\vec k\vec q}(t,t_0)$. Inserting the formal solutions for the functions
$U_{\vec q_1\vec q_2}(t,t_0)$, $U_{\vec r\vec q}(t,t_0)$ and $U_{\vec k\vec q}(t,t_0)$
into the differential equation for the function $U_{d\vec q}(t,t_0)$ one obtains the
derivative $\partial U_{d\vec q}(t,t_0)/\partial t$ expressed in terms of
$U_{dq}(t,t_0)$, $U_{\vec k\vec q}(t,t_0)$, $U_{\vec q_1\vec q_2}(t,t_0)$ and
$U_{\vec r\vec q}(t,t_0)$. Inserting again into this equation the formal solutions
for the functions $U_{\vec q_1\vec q_2}(t,t_0)$, $U_{\vec r\vec q}(t,t_0)$ and
$U_{\vec k\vec q}(t,t_0)$, and repeating this process again and again, one
obtains:
\begin{eqnarray}
 {\partial U_{d\vec q}(t,t_0) \over\partial t} &=& -i \tilde V_{d\vec q}(t) + (-i)^2 \int\limits^t_{t_0}
 dt_1 \sum_{\vec k_1} \tilde V_{d\vec k_1}(t) \tilde V_{\vec k_1\vec q}(t_1) \nonumber\\
&+& (-i)^3 \int\limits^t_{t_0} \int\limits^{t_1}_{t_0} dt_1 dt_2
  \left[\sum_{\vec q_1,\vec k_1} \tilde V_{d\vec q_1}(t) \tilde V_{\vec q_1\vec k_1}(t_1)
  \tilde V_{\vec k_1\vec q}(t_2) + \sum_{\vec r_1\vec k_1} \ldots \right]\nonumber\\
&+& (-i)^4 \int\limits^t_{t_0} \int\limits^{t_1}_{t_0} \int\limits^{t_2}_{t_0}
 dt_1 dt_2 dt_3 \left[\sum_{\vec k_1\vec q_1\vec k_2} \tilde V_{d\vec k_1}(t)
 \tilde V_{\vec k_1\vec q_1}(t_1) \tilde V_{\vec q_1\vec k_2}(t_2)
 \tilde V_{\vec k_2\vec q} (t_3)  \right. \nonumber \\
&&\left. \hspace{3cm} +\sum_{\vec k_1\vec r_1\vec k_2}\ldots  + \ldots \right] +
 \ldots  \nonumber\\
&+& (-i)^2 \int\limits^t_{t_0} dt_1 \sum_{\vec p=\vec k_1,\vec q_1,\vec r_1}
  \tilde V_{d\vec p}(t) \tilde V_{\vec p d}(t_1)
   U_{d\vec q} (t_1,t_0)  \nonumber \\
&+& (-i)^3 \int\limits^t_{t_0} \int\limits^{t_1}_{t_0} dt_1 dt_2
  \left[\sum_{\vec q_1,\vec k_1} \tilde V_{d\vec q_1}(t)  \tilde V_{\vec q_1\vec k_1}(t_1)
  \tilde V_{\vec k_1d}(t_2) \right.\nonumber\\
&+& \left.  \sum_{\vec r_1\vec k_1} \ldots + \sum_{\vec k_1\vec q_1}
   \ldots + \sum_{\vec k_1\vec r_1} \ldots \right] U_{d\vec q}(t_2,t_0) + \ldots
\end{eqnarray}
In the WBL approximation this equation is reduced to the following form:
\begin{equation}
 {\partial U_{d\vec q}(t,t_0) \over\partial t} = -C_4 \tilde V_{d\vec q}(t) + C_3 U_{d\vec q}
 (t,t_0) \,,
\end{equation}
where $C_4 = (1 - x)/( 1 + 2x^2)$ and has the solution as follows:
\begin{equation}
 U_{d\vec q}(t,t_0) = -C_4\int^t_{t_0} dt_1 \tilde V_{d\vec q}(t_1)
 {\rm exp} (-C^\star_3(t - t_1))\,.
\end{equation}
The function $U_{d\vec r}(t,t_0)$ is identical with $U_{d\vec q}(t,t_0)$.

To calculate, e.g. the current $j_L(t)$, we still need the functions $U_{\vec k \vec r}(t,t_0)$ and $U_{\vec k \vec q}(t,t_0)$.
The function $U_{\vec k\vec q}(t,t_0)$ can be obtained solving the set of the coupled differential equations for the functions
$U_{d\vec q}(t,t_0)$, $U_{\vec q\vec q_1}(t,t_0)$, $U_{\vec r\vec q}(t,t_0)$ and $U_{\vec k\vec q}(t,t_0)$. Writing down the
formal solution for $U_{\vec k\vec q}(t,t_0)$ and inserting into it, in the first step, the formal solutions for $U_{\vec
q_1\vec q_2}(t,t_0)$ and $U_{\vec r\vec q}(t,t_0)$ and, in the second step, the formal solutions for $U_{d\vec q}(t,t_0)$ and
$U_{\vec k\vec q}(t,t_0)$, and so on, one obtains
\begin{eqnarray}
\lefteqn{
U_{\vec k\vec q}(t,t_0) = -i\int\limits^t_{t_0} dt_1 \tilde V_{\vec kd}(t_1) U_{dq}
 (t_1,t_0)} \nonumber \\
&&  +(-i)^2\int\limits^t_{t_0} dt_1dt_2 \left[\sum_{\vec q_1} \tilde V_{\vec k\vec q_1}
 (t_1) \tilde V_{\vec q_1d}(t_2) + \sum_{\vec r_1} \tilde V_{\vec k\vec r_1}
 (t) \tilde V_{\vec r_1d}(t_2) \right] U_{d\vec q}(t_1,t_0) \nonumber\\
&+& (-i)^3 \int\limits^t_{t_0} \int\limits^{t_1}_{t_0}\int\limits^{t_2}_{t_0} dt_1 dt_2
  dt_3 \left[\sum_{\vec q_1,\vec k_1}
 \tilde V_{\vec kq_1}(t_1) \tilde V_{\vec q_1\vec k_1}(t_2) \tilde V_{\vec k_1d}(t_3)
+ \sum_{\vec r_1\vec k_1} \ldots \right] U_{d\vec q} (t_3,t_0) \nonumber \\
&+& \ldots + (-i) \int\limits^t_{t_0} dt_2 \tilde V_{\vec k\vec q} (t_1,t_0)
 + (-i)^3 \int\limits^t_{t_0} \int\limits^{t_1}_{t_0} \int\limits^{t_2}_{t_0}
  dt_1 dt_2 d_3\nonumber \\
&& \hspace{3cm} \left[\sum_{\vec q_1,\vec k_1}\tilde V_{\vec kq_1}(t_1) \tilde V_{\vec q_1\vec k_1}(t_2)
 \tilde V_{\vec k_1d}(t_3) + \sum_{\vec r_1\vec k_1} \ldots\right] + \ldots
\end{eqnarray}
Under the WBL approximation this equation reduces to the form:
\begin{equation}
 U_{\vec k\vec q} (t,t_0) = -C_2 \int\limits^t_{t_0} dt_1 \tilde V_{\vec kd}(t_1) U_{d\vec q}(t_1,t_0)
 - {i \over 1+2x^2} \int\limits^t_{t_0} dt_1\tilde V_{\vec k\vec q}(t_1)\,,
\end{equation}
where $U_{dq}(t,t_0)$ is given in Eq. (43). The function $U_{\vec k\vec r}(t,t_0)$ has the identical form as $U_{\vec k\vec
q}(t,t_0)$.


\end{document}